\documentclass[aps,twocolumn,
amsmath,amssymb]{revtex4-2}
\usepackage{color}
\usepackage{graphicx}
\usepackage{float}
\usepackage{subfigure}
\usepackage{booktabs}
\usepackage{dcolumn}
\usepackage{bm}
\usepackage{amsmath}
\usepackage{accents}
\usepackage{verbatim}
\usepackage{lineno}
\usepackage{tikz,pgfplots}
\usepackage{appendix}
\usepackage{braket}

\newcommand{\beq}{\begin{eqnarray}}
\newcommand{\eeq}{\end{eqnarray}}

\begin{document}
\title{Dynamical theory of the East model. I. A review on the  Zwanzig–Mori projection-operator formalism.}

\author{Chong Zha$^{1,2}$}
\author{Yuliang Jin$^{1,2,3}$}
\email{yuliangjin@mail.itp.ac.cn}


\affiliation{$^1$Institute of Theoretical Physics,
 Chinese Academy of Sciences, Beijing 100190, China}
\affiliation{$^2$School of Physical Sciences, University of Chinese Academy of Sciences, Beijing 100049, China}
\affiliation{$^3$Wenzhou Institute, University of Chinese Academy of Sciences, Wenzhou, Zhejiang 325000, China}

\date{\today}

\begin{abstract}
The East model is a paradigmatic kinetically constrained spin model in one dimension that exhibits glassy dynamics. This review surveys the progress on the theoretical description of the dynamics  of the East model, focusing on the spin time correlation function and the relaxation time. We summarize rigorous results on the super-Arrhenius scaling of the relaxation time in the low-temperature limit. We then present a unified derivation of the Zwanzig–Mori projection-operator formalism for both the fluctuation and domain bases. Within this formalism, an infinite hierarchy of coupled equations for the time correlation function and memory kernels emerges---a structure reminiscent of generalized mode-coupling theory developed  for supercooled liquids. We review the analytical approximations obtained by truncating or closing the hierarchy at first or second orders and compare them against Monte Carlo simulations. Although these approaches perform reasonably at moderate temperatures, they universally fail in the low-temperature regime: mode-coupling closure schemes in the fluctuation basis introduce spurious ergodicity-breaking transitions, while truncations in the domain basis discard slower modes and fail to capture the asymptotic divergence of the relaxation time. The loss of analytical tractability beyond second order remains the central obstacle, underscoring the fundamental difficulty of describing hierarchical relaxation even in this ``simplest'' model of glassy dynamics.
\end{abstract}
\maketitle

\section{Introduction}
Glassy dynamics refers to the  slow relaxation ubiquitous in disordered systems, including supercooled liquids~\cite{angell1995formation}, structural glasses~\cite{berthier2011theoretical}, spin glasses~\cite{binder1986spin,dahlberg2025spin}, and soft materials~\cite{cipelletti2005slow, ikeda2012unified}. Spin systems are prototypical models that exhibit glassy dynamics, yet they remain simple enough to have inspired sustained efforts toward exact theoretical solutions. Broadly speaking, glass-like spin models fall into two main categories.

(i) Spin glasses. Their glassy dynamics originate from quenched disorder, which is encoded in the spin couplings of the Hamiltonian. For mean-field spin glasses, rigorous results exist for the static structure of phase space and the thermodynamics~\cite{parisi1979infinite,parisi1980order}. On the dynamical side, dynamical mean-field theory yields a closed set of equations governing the two-time correlation and response functions~\cite{cugliandolo1993analytical}. However, despite being exact in infinite dimensions ($d\to\infty$), this approach fails to capture activated dynamics in low dimensions. Consequently, phenomenological descriptions—termed trap models—have been introduced to describe hopping between energy-landscape basins~\cite{bouchaud1997aging,bouchaud1995aging,li2026universal}. Nevertheless, a rigorous theory of dynamics for spin glasses in finite dimensions remains absent.

(ii) Kinetically constrained models (KCMs) \cite{Ritort2003Glassy}.
A distinctly different approach generates glassy dynamics through imposed kinetic constraints---i.e., dynamical rules that restrict local moves. 
A defining feature of KCMs is that they naturally embody cooperative facilitation: local relaxations can proceed only when neighboring sites are already mobile. 
Consequently, KCMs furnish an emergent framework for glassy dynamics that dispenses with explicit quenched disorder. 
Theoretical attention has been devoted to obtaining exact dynamical solutions for the one-dimensional ($d=1$) asymmetrically constrained kinetic Ising chain, known as the {\it East model}~\cite{jackle1991hierarchically}. 

As discussed above, spin-glass and KCM theories  approach the problem of glassy dynamics in finite dimensions from very different standpoints, differing in both their underlying mechanisms (quenched disorder versus dynamical rules) and their dimensional starting points ($d=\infty \to 3$ versus $d=1 \to 3$).

This article reviews the progress on the dynamical theory of the East model. The East model consists of a one-dimensional chain of binary (Ising) spins, $n_i \in \{0,1\}$, evolving according to Glauber dynamics. A kinetic constraint permits a spin to flip only if its right neighbor is in the excited state $n_{i+1}=1$. At temperature $T$, each spin is up with equilibrium probability 
\begin{equation}
c = \frac{1}{1+e^{1/T}}.
\label{eq:eq_c}
\end{equation}
Consequently, in the low-temperature limit ($T \to 0$ or $c \to 0$), excitations become sparse along the chain. Since these excitations are responsible for facilitating local relaxation, the dynamics slows down dramatically as the temperature decreases. 
The East model does not exhibit ergodicity breaking (an ergodic-nonergodic transition) at any finite temperature; instead, its dynamics freezes completely only at zero temperature. The most challenging regime for theoretical description is the low-temperature limit, where the dynamics becomes extremely slow, exhibits glass-like two-step relaxation, and yet remains fully ergodic.

In this paper, we focus on the equilibrium dynamics of the East model, typically characterized by the spin time correlation function (spin auto-correlation function) and the associated relaxation time (or correlation time). 
Rigorous results for the asymptotic scaling of the relaxation time in the low-temperature limit are available~\cite{aldous2002asymmetric,cancrini2008kinetically,cancrini2010facilitated,toninelli2012aging}:  the temperature-dependence of the relaxation time is super-Arrhenius, resembling typical behavior observed in fragile glass-forming liquids~\cite{angell1995formation}.

Unlike the scaling of the relaxation time, exact solutions for the low-temperature spin time correlation function remain elusive. Most existing theoretical approaches rely on the well-known Zwanzig–Mori projection operator formalism, in which the dynamics is projected onto the subspace of the chosen relevant variables~\cite{jackle1991hierarchically,pitts2000facilitated,pitts2001diagrammatic,eisinger1993analytical,wu2004east, kawasaki1995irreducible, van2005glassy}.  
The complementary, irrelevant part  feeds back into the evolution of the relevant variables indirectly, with this feedback fully encoded in a memory kernel that introduces a non-Markovian, history-dependent contribution. For any given complete basis, a hierarchy of coupled equations for the correlation function and memory kernels can be formally written, which is strikingly similar to the hierarchy in the generalized mode-coupling theory (GMCT)~\cite{mayer2006cooperativity, janssen2016generalized, luo2020generalized1, luo2020generalized2, biezemans2020glassy}. 
Unfortunately, solving this  hierarchy of equations remains impossible.

The central challenge thus becomes the design of a suitable basis, as this choice determines both the tractability and accuracy of the approximate memory kernel. The fluctuation basis is the standard choice~\cite{pitts2000facilitated,pitts2001diagrammatic,wu2004east}.  For this basis, a first-order truncation using the mode-coupling closure yields the well-known mode-coupling theory (MCT)~\cite{kawasaki1995irreducible}, which incorrectly predicts an ergodic-nonergodic transition at a finite temperature. More recently, a domain basis has been proposed~\cite{van2005glassy}. A first-order truncation, retaining only the single-domain basis, yields an explicit expression for the spin time correlation function~\cite{van2005glassy}. This approximation successfully avoids ergodicity breaking at all finite temperatures, but it neglects slow modes arising from higher-order domains. For both fluctuation and domain bases, analytical progress has been made by truncating or closing the hierarchy at second order. However, the convergence behavior of the solution remains unknown, as analytical tractability is lost beyond second order.

The rest of this paper is organized as follows. In Sec.~\ref{sec:model}, we describe the East model and the relevant dynamical observables. In Sec.~\ref{sec:time}, we review analytical results on the low-temperature scaling of the relaxation time. In Sec.~\ref{sec:Zwanzig-Mori}, we review the Zwanzig–Mori projection-operator formalism for the time correlation function, with detailed discussion of the general framework, its application to the fluctuation and domain bases, and mode-coupling approximations. These theoretical results are compared to Monte Carlo simulations in Sec.~\ref{sec:simulations}.
The hierarchical relaxation in microscopic dynamics is discussed in Sec.~\ref{sec:hierarchical}, and the paper is summarized 
 in Sec.~\ref{sec:discussion}.

\section{Model and dynamical observables}
\label{sec:model}

\subsection{Model}

The stochastic dynamics of the East model is governed by the master equation:
\begin{equation}
    \begin{split}
    \partial_tP({\bf n},t) 
    =\sum_i{[\tilde{\omega}_{\bar{n}_i}P({\bf n'},t)-\tilde{\omega}_{n_i}P({\bf n},t)]} = -\mathcal{L}P({\bf n},t),
    \end{split}
\label{eq:masterEQ}
\end{equation}
where  $\mathcal{L}$ is the Markov operator,  ${\bf n} = \{n_0, \ldots, n_i, \ldots, n_{N}\}$, ${\bf n'} = \{n_0, \ldots, \bar{n}_i, \ldots, n_{N}\}$, and $\bar{n}_i=1-n_i$. In this equation, we use 
\begin{equation}
\tilde{\omega}_{n_i}= \omega_{n_i} n_{i+1} = [c+(1-2c)n_i]n_{i+1}
\end{equation}
to denote the transition rate for the process $n_i\to \bar{n}_i$, and $P({\bf n},t)$ for the probability of the configuration ${\bf n}$. Note that,
\begin{equation}
\omega_{n_i} = c+(1-2c)n_i
\label{eq:omega}
\end{equation}
is the transition rate without taking into account facilitation.

The detailed balance condition gives a trivial equilibrium distribution
\begin{equation}
    P_{\rm eq}({\bf n})=\frac{1}{Z}e^{-\beta\sum_i n_i},
    \label{eq:EquilibriumDistribution}
\end{equation}
corresponding to the noninteracting Hamiltonian $H=\sum_i n_i$, where $\beta = 1/T$ is the inverse temperature, and $Z$ is the partition function. The factorized equilibrium distribution implies the absence of static correlations.

\subsection{Time correlation function}

The relaxation process of the East model in equilibrium is characterized by the auto-correlation function of single spin fluctuations: 
\begin{equation}
    \begin{split}
    C(t)&=\frac{\langle\delta n_i(t)\delta n_i(0)\rangle}{c(1-c)}\\
    &=\frac{P(n_i(t)=1|n_i(0)=1)-c}{1-c}
    ,
    \end{split}
    \label{eq:CorrFunc}
\end{equation}
where $\delta n_i=n_i-c$. The corresponding relaxation (correlation) time is defined as
\begin{equation}
    \tau=\int_0^{+\infty}{C(t)dt}.
\label{eq:CorrRelaxationTime}
\end{equation}
The correlation function $C(t)$ and time $\tau$ are the key objectives for the dynamical theory of the East model. Note that the cross-correlation between two different spins ($i \neq j$) is always zero, $\langle\delta n_i(t)\delta n_j(0)\rangle = 0$~\cite{pitts2000facilitated, van2005glassy}.

\section{Low-temperature scaling of the relaxation time}
\label{sec:time}

The mean relaxation time $\tau$ exhibits a super-Arrhenius divergence in the low temperature limit. The scaling behavior, 
\begin{equation}
\log{\tau}_{\rm energy} \sim \frac{(\log{1/c})^2}{\log{2}}, 
\label{eq:tau1}
\end{equation}
was first identified by Eisinger and Jäckle \cite{eisinger1993analytical}. A heuristic derivation based on domain coarsening was later developed by Sollich and Evans, showing that this leading scaling originates from the hierarchical energy-barrier structure of the East model~\cite{sollich1999glassy, sollich2003glassy}. 

The argument given by Sollich and Evans is insightful and can be summarized as follows. At low temperatures, the relaxation is dominated by {\it domains}. A domain is defined as a contiguous string of 0-spins between two 1-spins, together with the one at the right end.
 For example, the configuration $\ldots1|0001|\ldots$ represents a domain of length $4$. Note that $\ldots1|1|\ldots$corresponds to a domain of length $1$. It can be shown that the minimal barrier height $h(d)$ to relax a domain of size $d$ is, 
 \begin{equation}
 h(d) \sim \log_2 d.
 \label{eq:domainhierarchy}
 \end{equation}
 Indeed, in order to relax a domain of size $d$, the lowest cost is to flip the 0-spin in the middle, and then use it as an ``anchor'' to relax the two parts (each of size $d/2$) independently. Applying this procedure recursively then gives  Eq.~(\ref{eq:domainhierarchy}). 
 The relaxation time due to this ``energetic'' effect is 
 \begin{equation}
 \log{\tau}_{\rm energy} \sim \frac{h(d)}{T}.
 \label{eq:tau_energy_relation}
 \end{equation}
At low temperatures, the average domain size is~\cite{sollich1999glassy} 
\begin{equation}
    \langle d\rangle\sim 1/c.
    \label{eq:domainLengthEq}
\end{equation} 
Substituting the equilibrium domain size \(\langle d\rangle\) into
Eq.~(\ref{eq:tau_energy_relation}) yields Eq.~(\ref{eq:tau1}), since
 $\tau_{\rm energy} \sim e^{\frac{h(\langle d\rangle)}{T}} = e^{\frac{1}{T^2 \log 2}} \sim  e^{\frac{(\log{1/c})^2}{\log{2}}}$.

More recently, Cancrini et al. provided a rigorous analysis of the spectral gap, showing that the correct asymptotic scaling takes the form 
\begin{equation}
\log{\tau}\sim \frac{(\log{1/c})^2}{2\log{2}},
\label{eq:tau}
\end{equation}
thereby refining the earlier prediction by identifying an additional factor of $1/2$ in the exponent \cite{cancrini2008kinetically}. 

While the leading scaling can be understood from energetic considerations alone, the additional factor $1/2$ in the exponent reflects the contribution of the multiplicity of the dynamical pathways. By means of graph-theoretic and combinatorial arguments, Chung et al. derived an estimate for the number of configurations reachable from the global ground state through trajectories whose maximal barrier height is $h$ \cite{chung2001combinatorics}. Denoting this number by $V(h)$, they found $V(h)\sim \xi^h h!2^{h(h-1)/2}$, where $0<\xi<1$ is a subleading factor that may depend on $h$, and does not modify the leading asymptotic scaling behavior. Building on this result, Cancrini et al. provided a more natural derivation of the relaxation-time scaling \cite{cancrini2010facilitated}. 
The quantity $V(h)$ is independent of $T$. 
If one defines an ``entropy'' by $s = \log{V(h)}$, then to leading order:
\begin{equation}
s(d)\approx \log{2^{h^2(d)/2}}=\frac{h^2(d)\log{2}}{2}.
\label{eq:entropy}
\end{equation}
Taking into account both the energetic (Eq.~\ref{eq:tau1}) and entropic (Eq.~\ref{eq:entropy}) effects, the relaxation time becomes
\begin{equation}
\log \tau \sim \frac{F[h(d)]}{T},
\label{eq:tau_total}
\end{equation}
where the effective free energy of a given domain length $d$ is defined as,
\begin{equation}
F(d) = h(d)- T s(d).
\label{eq:free-energy}
\end{equation} 
The equilibrium relaxation time is then
$\tau \sim e^{\frac{h(\langle d\rangle)}{T} - s(\langle d\rangle)}  = \frac{1}{2T^2\log{2}}$, where we have used $h(\langle d\rangle) \sim \log_2 \langle d\rangle$ (see Eq.~\ref{eq:domainhierarchy}) and $\langle d\rangle \sim 1/c\sim e^{1/T}$ (see Eq.~\ref{eq:domainLengthEq}). This derivation gives Eq. (\ref{eq:tau}).

The asymptotic scaling of $\tau$ reveals two important points. First, the temperature dependence of the relaxation time is super-Arrhenius, with $\tau \sim e^{\frac{1}{2 T^2 \log 2}}$---a hallmark of fragile glasses. Second, $\tau$ diverges as $T \to 0$, implying the absence of ergodic-nonergodic transition at any finite temperature.

\section{Zwanzig–Mori projection-operator formalism for the time correlation function}
\label{sec:Zwanzig-Mori}

\subsection{Zwanzig-Mori formalism}
The Zwanzig-Mori formalism describes stochastic dynamics by projecting the time-evolution operator onto subspaces of the Hilbert space spanned by dynamical variables~\cite{mori1965transport,zwanzig1960ensemble}. Consider a dynamical variable, $A(t)$, and its time-correlation function, $C(t) = \langle A(t) A(0) \rangle$. The dynamics of $C(t)$ follows the general equation:
\begin{equation}
\frac{d}{dt} C(t) =  -\Omega_0 C(t) + \int_0^t d\tau M(t-\tau) C(\tau),
\label{eq:corr_real}
\end{equation}
where $\Omega_0$ is the frequency of the instantaneous, memory-less dynamics, and $M(\tau)$ is the memory kernel.

The phase space of the equilibrium ensemble, $\{ \bf{n} \}$, has a dimensionality $2^N$. Any dynamical variable can be expanded for a chosen complete orthonormal static basis $\{|e_k\rangle\}$: 
\begin{equation}
    |A(t) \rangle=\sum_k|e_k\rangle\langle e_k|A(t)\rangle,
    \label{eq:expansionOff}
\end{equation}
where we have used the vector notation $|A(t) \rangle$ to denote the representation of $A(t)$ in the given basis. The inner product is defined with respect to an equilibrium probability measure $P_{\rm eq}({\bf n})$ as
\begin{equation}
    \langle A|B\rangle:=\sum_{{\bf n}}{A({\bf n})B({\bf n})P_{\rm eq}({\bf n})}.
    \label{eq:innerProd}
\end{equation}

For our model, $|A(t) \rangle$ is the normalized single spin fluctuation (see Eq.~\ref{eq:CorrFunc}), which naturally determines the first basis vector:
\begin{equation}
    |A(0) \rangle = |e_0\rangle = \frac{n_0-c}{\sqrt{c(1-c)}}.
    \label{eq:theFirstBasisOfEast}
\end{equation}
Note that $\langle\delta n_i(t)\delta n_i(0)\rangle = \langle\delta n_0(t)\delta n_0(0)\rangle$ for any $i$.

The key idea of the Zwanzig-Mori formalism is to project  $|\dot{A}(t) \rangle$  onto the sub-space of $|e_0\rangle$ and its complementary-space, using operators,
\begin{equation}
    \begin{split}
        &\mathcal{P} = |e_0\rangle\langle e_0|,\\
        &\mathcal{Q} = \mathbf{I} - \mathcal{P} = \sum_{k\geq 1}|e_k\rangle\langle e_k|.
    \end{split}
    \label{eq:ProjectionOperatorExpressOnBasis}
\end{equation}
The projection $\mathcal{P}$ into the sub-space of $|e_0\rangle$ gives the term $-\Omega_0 C(t)$ in Eq.~(\ref{eq:corr_real}),  corresponding to a simple exponential decay with a rate $\Omega_0=-\langle e_0|\mathcal{L}|e_0\rangle = c$. Using the projection operators, the memory kernel can be expressed  as,
\begin{equation}
 \begin{split}
M(t) & = \langle \mathcal{Q} \mathcal{L} e_0 |  e^{\mathcal{Q} \mathcal{L} \mathcal{Q}t} | \mathcal{Q} \mathcal{L} e_0 \rangle \\
& = \sum_{k,k'\geq 1} L_{0k} \langle e_k | e^{\mathcal{Q}\mathcal{L}\mathcal{Q} t} | e_{k'} \rangle L_{k'0}
 \end{split}
 \label{eq:Q_real}
\end{equation}
where Eq.~(\ref{eq:ProjectionOperatorExpressOnBasis}) has been used, and $L_{kk'} = \langle e_k|\mathcal{L}|e_{k'}\rangle$. 

In practice, it is often more convenient to treat the  Laplace transform of the time-correlation function, $\tilde{C}(z) = \int_0^\infty dt e^{-zt} C(t)$. In the Laplace space, 
Eqs.~(\ref{eq:corr_real}) and ~(\ref{eq:Q_real}) become,
\begin{equation}
\begin{split}
    \tilde{C}(z) & = \langle e_0| \tilde{R}(z) |e_0\rangle \\
    & = \frac{1}{z+\Omega_0-\tilde{M}(z)},
\end{split}    
    \label{eq:FirstOrderZwanzig}
\end{equation}
where
\begin{equation}
    \tilde{R}(z)=[z-\mathcal{L}]^{-1}
    \label{eq:Resolvent}
\end{equation}
is the resolvent,
and 
\begin{equation}
    \tilde{M}(z)=\sum_{k,k'\geq 1} L_{0k}\langle e_k| \tilde{R}_{\mathcal{Q}}(z)|e_{k'}\rangle L_{k'0},
    \label{eq:FirstMemoExpression}
\end{equation}
where 
\begin{equation}
    \tilde{R}_{\mathcal{Q}}(z)=[z-\mathcal{Q}\mathcal{L}\mathcal{Q}]^{-1}
    \label{eq:complementaryResolvent}
\end{equation}
 is the resolvent of the complementary space. 

\subsection{Hierarchy of coupled
equations}
\label{se:GeneralHierarchyEq}

A distinctive feature of the East model is that the kinetic constraint is strictly local: each spin interacts dynamically only with its nearest neighbor. As a consequence, it is always possible to organize the orthonormal basis into a sequence of basis sets,
\begin{equation}
    |{\bf q}_0\rangle=\{|e_0\rangle\},|{\bf q}_1\rangle,|{\bf q}_2\rangle,...,
    \label{eq:DefbasisSet}
\end{equation}
such that the Markov operator $\mathcal{L}$ couples only neighboring basis sets between $|{\bf q}_k\rangle$ and $|{\bf q}_{k+1}\rangle$. Note that $|{\bf q}_k\rangle$ generally contains multiple basis vectors. 

In this representation, the matrix representation ${\bf L}$ of the Markov operator $\mathcal{L}$ has  a block-tridiagonal form,
\begin{equation}
    {\bf L}=\begin{pmatrix}
        {\bf L}_{00} & {\bf L}_{01} & 0 & ... \\
         {\bf L}_{10} & {\bf L}_{11} & {\bf L}_{12} & ... \\
         0 & {\bf L}_{21} & {\bf L}_{22} & \ddots \\
         \vdots & \vdots & \ddots & \ddots 
    \end{pmatrix},
    \label{eq:MarkovBlockMatrix}
\end{equation}
where each element ${\bf L}_{k k'} =  \langle {\bf q}_k|\mathcal{L} |{\bf q}_{k'} \rangle$ is generally a matrix.

Consequently, when the basis sets rather than the individual basis vectors are regarded as the elementary building blocks, the summation appearing in Eq. (\ref{eq:FirstMemoExpression}) reduces to a single non-vanishing contribution
\begin{equation}
     \tilde{M}(z) = {\bf L}_{01}\langle {\bf q}_1|R_{\mathcal{Q}}(z)|{\bf q}_1\rangle {\bf L}_{10} = {\bf L}_{01}      \tilde{
\bf C}_1^r(z) {\bf L}_{10},
    \label{eq:OnlyTermInMemo}
\end{equation}
where
\begin{equation}
    \tilde{
\bf C}_1^{\rm r}(z)=\langle {\bf q}_1|\tilde{R}_{\mathcal{Q}_1}(z)|{\bf q}_1\rangle
\label{eq:C1r}
\end{equation}
is a reduced time-correlation function.
All coupling to more distant basis sets vanish identically: ${\bf L}_{0k}=0$ for $k>1$.

It is worth noting that $\tilde{
\bf C}_1^{\rm r}(z)$ in Eq.~(\ref{eq:C1r})
has exactly the same mathematical structure as the correlation function itself (see Eq.~\eqref{eq:FirstOrderZwanzig}): both are given by the leading matrix element of the resolvent operator in an appropriate linear space. Moreover, the restriction of the Markov operator to the complementary subspace retains the same block-tridiagonal structure. Consequently, the projection-operator decomposition can be applied recursively to the complementary subspace, generating an infinite hierarchy of coupled equations (when $N \to \infty$), or equivalently, an infinite matrix continued fraction.

To formalize this hierarchy, we introduce the $k$-th complementary-space projector:
\begin{equation}
    \mathcal{Q}_k=\sum_{j\geq k}|{\bf q}_j\rangle\langle {\bf q}_j|,k=0,1,...,
    \label{eq:DefOfkOrderQ}
\end{equation}
and define the corresponding reduced correlation function
\begin{equation}
    \tilde{\bf C}_k^{\rm r}(z)=\langle {\bf q}_k|\tilde{R}_{\mathcal{Q}_k}(z)|{\bf q}_k\rangle,
    \label{eq:Def_kOrderReducedCorrFunc}
\end{equation}
The resulting hierarchy is then given by
\begin{equation}
    \tilde{\bf C}_k^{\rm r}(z)=[z {\bf I}- {\bf L}_{kk}-{\bf L}_{k,k+1} \tilde{\bf C}_{k+1}^{\rm r}(z) {\bf L}_{k+1,k}]^{-1}.
    \label{eq:k-thEq}
\end{equation}
for $k=0, 1, \ldots$, with the generalized $k$-th order memory kernel, 
\begin{equation}
\tilde{\bf M}_k(z)={\bf L}_{k,k+1}\tilde{\bf C}_{k+1}^{\rm r}(z){\bf L}_{k+1,k}.
\end{equation}
Note that by definition, ${\tilde{\bf C}}_0^{\rm r}(z)=\tilde{C}(z)$, and the $(k=0)$-th order equation is identical to the original equation of the correlation function, Eq.~(\ref{eq:FirstOrderZwanzig}).

The hierarchy in Eq.~(\ref{eq:k-thEq}) is remarkable. It shows that, in the thermodynamic limit $N\to \infty$, the time-correlation function of the East model is determined by an infinite number of coupled equations. In the zero-temperature limit, truncation at any level inevitably introduces approximations. This explains why the analytical treatment of the dynamics of the East model is so difficult: to date, no exact solution has been developed for this one-dimensional model. Typically, approximations need to be introduced either by truncating the hierarchy (setting $\tilde{\bf C}_k^{\rm r}(z) = {\bf 0}$ at the $k$-th level), or by imposing closures such as $\tilde{\bf C}_k^{\rm r}(z) = \tilde{\bf C}_{k-1}^{\rm r}(z)$. Such treatments will be discussed in the next two subsections for the fluctuation and domain bases.

The hierarchical structure of the East model bears a striking resemblance to the recently formulated generalized mode-coupling theory (GMCT) for supercooled liquids~\cite{mayer2006cooperativity, janssen2016generalized, luo2020generalized1, luo2020generalized2, biezemans2020glassy}, which similarly involves an infinite number of hierarchically coupled equations. This observation reveals the mathematical origin of the dynamical analogy between such a simple one-dimensional system and the relaxation of supercooled liquids. 
Generally, the absence of a typical scale  in the infinite hierarchy gives rise to a non-Arrhenius slowdown with a decreasing temperature---in contrast to Arrhenius behavior, which is governed by a single activation energy. In addition, the infinite hierarchy inherently prevents ergodic-nonergodic transition at any finite temperature. An apparent finite-temperature ergodic-nonergodic transition would emerge only if the hierarchy were artificially truncated via closure approximations; such cases will be discussed in Sec.~\ref{sec:MCT} within the MCT framework.
A crucial difference between Eq.~(\ref{eq:k-thEq}) and GMCT lies in the nature of the reduced correlation functions: while GMCT deals with scalars, Eq.~(\ref{eq:k-thEq}) involves the matrix $\tilde{\bf C}_k^{\rm r}(z)$. Consequently, the Z-transform technique established for the schematic GMCT~\cite{janssen2016generalized} cannot be directly transferred to the current problem.

The feasibility of the hierarchy is critically dependent on the choice of complete basis. Below, we will discuss two complete bases introduced in prior work~\cite{wu2004east, van2005glassy}. For both bases, the problem remains analytically intractable beyond second order. Although the original theories are derived in a different manner, we present them below within a unified framework, as outlined in this subsection.

\subsection{Irreducible memory kernels}
\label{se:irreducible}

In the East model, direct approximations to the memory kernel do not lead to useful dynamical predictions~\cite{eisinger1993analytical}. Kawasaki introduced the concept of the irreducible memory kernel \cite{kawasaki1995irreducible}, which is related to the conventional memory kernel through
\begin{equation}
    \tilde{M}(z)=\frac{\tilde{M}^{\rm ir}(z)}{1+\Omega_0^{-1}\tilde{M}^{\rm ir}(z)}.
    \label{eq:relationofMandMir}
\end{equation}
The irreducible memory kernel isolates the part of the dynamics that cannot be generated through repeated propagation within the projected subspace. Consequently, approximate closures imposed on the irreducible kernel are generally more accurate than direct approximations to the conventional memory kernel. 

Wu and Cao extended this idea by introducing a recursive definition of irreducible dynamics at arbitrary projection levels~\cite{wu2004east}. 
In this irreducible representation, the  hierarchy Eq.~(\ref{eq:k-thEq}) becomes,
\begin{equation}
\begin{split}
    &\tilde{\bf C}_k^{\rm ir}(z)=\\
   & \left[z {\bf I} + \delta {\bf \Omega}_k \left( \delta {\bf \Omega}_k  + {\bf L}_{k,k+1} \tilde{\bf C}_{k+1}^{\rm ir}(z) {\bf L}_{k+1,k} \right)^{-1} \delta {\bf \Omega}_k  \right]^{-1}, 
    \end{split}
    \label{eq:k-thEq_irr}
\end{equation}
Here, $\delta {\bf \Omega}_k = {\bf \Omega}_k - {\bf \Omega}_k^L(0)$, with ${\bf \Omega}_k  = {\bf L}_{kk}$ and ${\bf \Omega}_k^L(z):=[\tilde{\bf C}_k^{r}(z)]^{-1}-[\tilde{\bf C}_k(z)]^{-1}$, where   $\tilde{\bf C}_k(z):=\langle {\bf q}_k|\tilde{R}(z)|{\bf q}_k\rangle$. 
The frequency matrix ${\bf \Omega}_k^L(z)$ can be solved 
from the following recursion relation,
\begin{equation}
    {\bf \Omega}_k^L(z)={\bf L}_{k,k-1}[z{\bf I}-{\bf L}_{k-1,k-1}-{\bf \Omega}_{k-1}^L(z)]^{-1}{\bf L}_{k-1,k},
    \label{eq:RecuOmegaL}
\end{equation}
with the initial condition ${ \Omega}_0^L(z)=0$.  
Thus, $\delta {\bf \Omega}_k$ is known for the given matrix ${\bf L}$. 
Once $\tilde{\bf C}_k^{\rm ir}(z)$ is solved from Eq.~(\ref{eq:k-thEq_irr}), the reduced correlation function can be obtained from the relation, 
$[\tilde{\bf C}_k^{\rm ir}(z)]^{-1}=[\tilde{\bf C}_k^{\rm r}(z)]^{-1}-{\bf \Omega}_k^L(0)$.

\subsection{Fluctuation basis and closures up to the second order}
\label{sec:fluctuation}
The fluctuation basis consists of products of single-spin fluctuation variables and forms a complete orthonormal basis after normalization~\cite{pitts2001diagrammatic},
\begin{equation}
    A_m(i_1i_2...i_m)=\prod_{j=1}^mA_1(i_j),
    \label{eq:Def_fluctuationBasis}
\end{equation}
where $0 \leq i_j \leq N$ is the spin index and ordered as $i_1 < i_2 < ... < i_m$.
The single spin fluctuation variable is defined as $A_1(i)=(n_i-c)/\sqrt{c(1-c)}$.

For a one-dimensional chain of length $N$, the total number of fluctuation basis functions is $\sum_{k=1}^N\binom{N}{k}=2^N-1$. The remaining basis function is the constant mode, which is not included since it does not represent a fluctuation. However, only those basis functions generated from fluctuations containing the leftmost spin are dynamically coupled to the single-spin fluctuation $A_1(0)$. Consequently, it is sufficient to consider half of the space: the $2^{N-1}$ basis functions originating from $A_1(0)$~\cite{van2005glassy} (the other half does not contain $A_1(0)$). This shows that the $2^{N-1}$ basis in Eq.~(\ref{eq:Def_fluctuationBasis}) is complete for the dynamics of the East model. 

Based on the  fluctuation basis, Pitts and Anderson originally developed a theory using a diagrammatic expansion rather than the Zwanzig-Mori projection-operator formalism~\cite{pitts2001diagrammatic}. Subsequently, Wu and Cao reformulated the theory within the Zwanzig-Mori framework and demonstrated that the resulting hierarchical expansion is mathematically equivalent to the original diagrammatic construction of Pitts and Anderson~\cite{wu2004east}.

In the fluctuation basis, Wu and Cao introduced a natural hierarchical decomposition of the basis into a sequence of basis sets. Starting from $|{\bf q}_0\rangle=|A_1(0)\rangle$, the next basis set $|{\bf q}_1\rangle$ is defined as the collection of all basis functions that are directly coupled to $|{\bf q}_0\rangle$ through the action of Markov operator $\mathcal{L}$. Higher-order basis sets are then generated recursively in the same manner, with $|{\bf q}_{k+1}\rangle$ consisting of the basis functions newly reached by applying the Markov operator to $|{\bf q}_k\rangle$. Under this recursive construction, the Markov operator naturally leads to a block tridiagonal structure, so that the hierarchical continued-fraction equations of Eq. (\ref{eq:k-thEq}) follows directly. The price paid for this construction is that the dimension of each basis set grows combinatorially with every recursive step. As a consequence, obtaining an exact analytical solution of the resulting infinite hierarchy rapidly becomes intractable, making approximation closure schemes at low orders the only practical route.

In the fluctuation basis, the first three matrix irreducible memory kernels reduce to scalars, $\tilde{C}^{\rm ir}(z)$, $\tilde{C}_1^{\rm ir}(z)$, $\tilde{C}_2^{\rm ir}(z)$. 
The first equation of the hierarchy Eq.~(\ref{eq:k-thEq_irr}) is
\begin{equation}
    \tilde{C}(z)=\frac{1}{z+c-\frac{c(1-c)\tilde{C}_1^{\rm ir}(z)}{1+\Omega_1^L(0) \tilde{C}_1^{\rm ir}(z)}},
    \label{eq:1stEq}
\end{equation}
where $\Omega_1^L(0)=-L_{10}L_{01}/L_{00}=1-c$, according to Eq. (\ref{eq:RecuOmegaL}). In the first-order simple mode-coupling (SMC) closure, one uses the approximation,
\begin{equation}
    \tilde{C}_1^{\rm ir}(z)\approx\tilde{C}(z),
    \label{eq:1stSMC}
\end{equation}
which leads to an explicit expression for $\tilde{C}(z)$. This approximation provides good agreement with numerical simulations for $c>0.5$. However, it incorrectly predicts an ergodicity-breaking transition at $c=0.5$.

The second-order equation is 
\begin{equation}
    \tilde{C}_1^{\rm ir}(z)=\frac{1}{z+c-\frac{L_{12}L_{21}\tilde{C}_2^{\rm ir}(z)}{1+\Omega_2^L(0) \tilde{C}_2^{\rm ir}(z)}}.
    \label{eq:Cir1Expression}
\end{equation}
The second-order SMC closure,
\begin{equation}
    \tilde{C}_2^{\rm ir}(z)\approx \tilde{C}_1(z),
    \label{eq:2ndSMC}
\end{equation}
 terminates the hierarchy at the second recursion level.
However, this second-order closure does not resolve the fundamental deficiency of the first-order SMC. Although it reduces the height of the spurious non-ergodic plateau, the ergodicity-breaking transition at $c=0.5$ remains unchanged.

From a practical viewpoint, the second-order SMC represents the end of this construction. Beyond the second recursion level, the reduced correlation functions become matrix-valued rather than scalar-valued, and the dimensions of the memory kernels grow rapidly. Consequently, the simple scalar closure is no longer well defined, preventing a straightforward extension of the higher-order SMC hierarchy.

\subsection{Domain basis and truncations up to the two-domain basis}
\label{sec:domain}

The domain basis is introduced by van Zon and Schofield as an alternative complete orthonormal basis for the dynamical Hilbert space \cite{van2005glassy}. The first basis function is again chosen as the normalized single-spin fluctuation, while the higher-order basis functions are constructed by considering fluctuations associated with increasing numbers of domains, leading to a hierarchy classified according to the domain number. The $\alpha$-domain basis functions are defined as
\begin{equation}
    \begin{split}
        Q_\alpha(k_1,...,k_\alpha)=&\frac{n_0-c}{Z_\alpha(k_1,...,k_\alpha)}D_1(k_1)n_{k_1+1}D_{k_1+2}(k_2)\\
        &\times n_{k_1+k_2+2}...D_{k_1+...+k_{\alpha-1}+\alpha}(k_{\alpha})\\
        &\times (n_{k_1+...+k_\alpha+\alpha}-c),
    \end{split}
    \label{eq:Def_domainbasis}
\end{equation}
where $D_i(k_i):=\prod_{j=i}^{i+k_i-1}(1-n_j)$, $\alpha=0,1,2,\ldots$, $k_j=0,1,\ldots$ with $j=1,\ldots, \alpha$ and $D_i(0):=1$. The normalization constants satisfy
\begin{equation}
    Z_\alpha(k_1,...,k_\alpha)=\begin{cases}
        \sqrt{c(1-c)},\alpha=0,\\
        c^{(1+\alpha)/2}(1-c)^{1+\sum_{j=1}^{\alpha}k_j/2},\text{otherwise}.
    \end{cases}
    \label{eq:NormalizationOfDomainBasis}
\end{equation}
The parameter $\alpha$ denotes the number of domains, and $k_i$ denotes the length of the $i$-th domain.

In the original work, van Zon and Schofield progressively enlarged the projection space and formulated the resulting recursion relation in terms of the inversion of tridiagonal matrices. Here we recast their construction into the general framework introduced in \ref{se:GeneralHierarchyEq}. This reformulation allows the domain-basis approach to be directly compared with the fluctuation-basis hierarchy and, more importantly, provides a unified language for discussing different analytical approximations.

In this classification, the $m$-domain basis functions constitute the basis set ${\bf q}_m$. Owing to the nearest-neighbor kinetic constraint of the East model, the Markov operator couples only neighboring basis sets, so that its matrix representation again has a block-tridiagonal structure. For an infinite chain, each basis set ${\bf q}_m$ with $m\geq 1$ contains infinitely many basis functions. Nevertheless, owing to the special structure of the domain basis, most matrix elements within each block vanish identically. As a consequence, each diagonal block ${\bf L}_{mm}$ is itself a tridiagonal matrix. Moreover, up to the two-domain basis level, the corresponding memory kernels also preserve the tridiagonal structure. This ubiquitous tridiagonal structure allows the projection-operator formalism to be further applied within each individual basis set, enabling an analytical treatment of hierarchy and the calculation of the correlation function $\tilde{C}(z)$ with truncation up to the two-domain basis level.

The first nontrivial approximation is obtained by truncating the hierarchy after the first two basis sets, ${\bf q}_0$ and ${\bf q}_1$. This is equivalent to setting $\tilde{\bf C}_2^{\rm r}(z)={\bf 0}$, so that the infinite hierarchy Eq.~(\ref{eq:k-thEq}) reduces to the following two coupled equations
\begin{equation}
    \begin{split}
        &\tilde{C}(z)=[z- L_{00}-{\bf L}_{01} \tilde{\bf C}_{1}^{\rm r}(z) {\bf L}_{10}]^{-1},\\
        &\tilde{\bf C}_1^{\rm r}(z)=[z{\bf I}-{\bf L}_{11}]^{-1}.
    \end{split}
    \label{eq:1domainEq}
\end{equation}
where
\begin{equation}
    {\bf L}_{01}=\begin{pmatrix}
        -\sqrt{c(1-c)}&0&0&...
    \end{pmatrix},
    \label{eq:matrixL01}
\end{equation}
and
\begin{equation}
    {\bf L}_{11}=\begin{pmatrix}
        -1 & c\sqrt{1-c} & &  \\
        c\sqrt{1-c} & -c(2-c) & c\sqrt{1-c} &  \\
         & c\sqrt{1-c} & -c(2-c) & \ddots \\
         &  &\ddots &\ddots
    \end{pmatrix},
    \label{eq:matrixL11}
\end{equation}
with the omitted elements  repetitions of the same structure. Equation~(\ref{eq:matrixL01}) indicates that $\tilde{C}_0^{\rm r}(z) = \tilde{C}(z)$ couples only to the first diagonal element of $\tilde{\bf C}_1^{\rm r}(z)$. Therefore, the problem is reduced to evaluating the resolvent of a simple tridiagonal matrix. By applying the projection-operator formalism, one obtains a continued fraction,
\begin{equation}
    \begin{split}
       & [\tilde{\bf C}_1^{\rm r}(z)]_{00} \\
       &=\frac{1}{z+1-\frac{c^2(1-c)}{z+c(2-c)-\frac{c^2(1-c)}{\ddots}}}\\
        &=\frac{2}{z+2-c(2-c)+\sqrt{4cz+(z-c^2)^2}}.
    \end{split}
    \label{eq:dominBasisC1rExpression}
\end{equation}
Combining Eq. (\ref{eq:dominBasisC1rExpression}) and Eq. (\ref{eq:1domainEq}) gives the explicit formula of $\tilde{C}(z)$.

Similarly, one may truncate the hierarchy at the two-domain level by setting $\tilde{\bf C}_{3}^{\rm r}(z)=0$, thereby retaining only the $0$-, $1$- and $2$-domain basis sets. The hierarchical equations are then reduced to
\begin{equation}
    \begin{split}
        &\tilde{C}(z)=[z- L_{00}-{\bf L}_{01} \tilde{\bf C}_{1}^{\rm r}(z) {\bf L}_{10}]^{-1},\\
        &\tilde{\bf C}_1^{\rm r}(z)=[z{\bf I}-{\bf L}_{11}-{\bf L}_{12}\tilde{\bf C}_{2}^{\rm r}(z){\bf L}_{21}]^{-1},\\
        &\tilde{\bf C}_2^{\rm r}(z)=[z{\bf I}-{\bf L}_{22}]^{-1}.
    \end{split}
    \label{eq: 2domainEq}
\end{equation}
The key observation is that the tridiagonal structure is established recursively throughout the hierarchy. As discussed above, the special single-link structure of ${\bf L}_{01}$ implies that $\tilde{C}(z)$ couples only to the first diagonal element of $\tilde{\bf C}_1^{\rm r}(z)$. In the expression for $\tilde{\bf C}_1^{\rm r}(z)$, the block ${\bf L}_{11}$ is itself tridiagonal. 

The next step is to examine the coupling between the one- and two-domain basis sets. The only nonzero matrix elements of ${\bf L}_{12}$ are 
$\langle Q_1(k)|\mathcal{L} | Q_2(k-1,0) \rangle$ and $\langle Q_1(k)| \mathcal{L} |Q_2(k,0) \rangle$
for $k>0$, together with $\langle Q_1(0)| \mathcal{L}|Q_2(0,0) \rangle$. More importantly, the matrix ${\bf L}_{22}$, acting within the two-domain basis, possesses a partially separable structure. Since the basis functions $Q_2(k_1,k_2)$ are labeled by two indices, each matrix element carries two pairs of indices. The matrix ${\bf L}_{22}$ is diagonal with respect to the first index $k_1$, while it is tridiagonal with respect to the second index $k_2$. Consequently, its resolvent $\tilde{\bf C}_2^{\rm r}(z)$ remains diagonal in the first index.

The property immediately determines the structure of the memory kernel appearing in the equation for $\tilde{\bf C}_1^{\rm r}(z)$. Starting from a basis function $Q_1(k)$ with $k>0$, the matrix ${\bf L}_{12}$ connects only to $Q_2(k-1,0)$ and $Q_2(k,0)$. Propagation through $\tilde{\bf C}_2^{\rm r}(z)$ preserves the first index and therefore reaches only basis functions of the form $Q_2(k-1,m)$ and $Q_2(k,m')$. Since the coupling back through ${\bf L}_{21}$ is nonzero only for $m=m'=0$, the resulting memory kernel connects $Q_1(k-1),Q_1(k)$ and $Q_1(k+1)$. The effective memory kernel is itself tridiagonal.

As a result, the first equation of Eq. (\ref{eq: 2domainEq}) reduces to the inversion of a scalar function, whereas the second and third equations require only the first diagonal elements of the inversion of tridiagonal matrices. Furthermore, these tridiagonal matrices exhibit the same asymptotically repetitive structure as in the one-domain truncation: beyond a finite number of rows, the diagonal and off-diagonal entries become constant. Consequently, they can be solved analytically by the same continued-fraction technique.

The truncation at the two-domain level 
captures the long-time relaxation for temperatures down to $c\geq 0.3$. At lower temperatures, however, the approximation gradually deteriorates, with the predicted relaxation time incorrectly diverging as $\tau^{(2)}_{\rm DB} \sim c^{-4}$ \cite{van2005glassy}.

A central question is whether the same analytical strategy in Ref.~\cite{van2005glassy} can be extended by retaining the three-domain basis. Unfortunately, this turns out not to be the case. Truncating the hierarchy after the three-domain basis leads to the coupled equations
\begin{equation}
    \begin{split}
        &\tilde{C}(z)=[z- L_{00}-{\bf L}_{01} \tilde{\bf C}_{1}^{\rm r}(z) {\bf L}_{10}]^{-1},\\
        &\tilde{\bf C}_1^{\rm r}(z)=[z{\bf I}-{\bf L}_{11}-{\bf L}_{12}\tilde{\bf C}_{2}^{\rm r}(z){\bf L}_{21}]^{-1},\\
        &\tilde{\bf C}_2^{\rm r}(z)=[z{\bf I}-{\bf L}_{22}-{\bf L}_{23}\tilde{\bf C}_3^{\rm r}(z){\bf L}_{32}]^{-1},\\
        &\tilde{\bf C}_{3}^{\rm r}(z)=[z{\bf I}-{\bf L}_{33}]^{-1},
    \end{split}
    \label{eq: 3domainEq}
\end{equation}
where the block ${\bf L}_{33}$ acts on the basis functions $Q_3(k_1,k_2,k_3)$. Owing to the structure of the domain basis, ${\bf L}_{33}$ is diagonal with respect to the first two indices and tridiagonal with respect to the third one. Consequently, the corresponding reduced correlation function matrix $\tilde{\bf C}_3^{\rm r}(z)$ is also diagonal in the first two indices, and its leading matrix element can still be evaluated analytically using the projection-operator formalism.

The difficulty instead originates from the off-diagonal block ${\bf L}_{23}$. The nonzero matrix elements are $\langle Q_2(k_1,k_2)|\mathcal{L}|Q_3(k_1,k_2,0) \rangle$,$\langle Q_2(k_1,k_2)|\mathcal{L}|Q_3(k_1,k_2-1,0) \rangle$ and $\langle Q_2(k_1,k_2) |\mathcal{L}| Q_3(k_1-1,0,k_2) \rangle$. The first two couplings preserve the elegant tridiagonal structure encountered at the two-domain level. After propagation through $\tilde{\bf C}_3^{\rm r}(z)$, they connect only to basis functions of the form $Q_3(k_1,k_2,m)$ and $Q_3(k_1,k_2-1,m')$. Since the coupling back through ${\bf L}_{32}$ is nonzero only for $m=m'=0$, the resulting memory kernel connects only neighboring values of the second index, namely $Q_2(k_1,k_2-1),Q_2(k_1,k_2)$ and $Q_2(k_1,k_2+1)$. Were these the only contributions, the tridiagonal structure would survive unchanged.

The essential complication arises from the third coupling, 
$\langle Q_2(k_1,k_2) |\mathcal{L}| Q_3(k_1-1,0,k_2) \rangle$. Propagation through $\tilde{\bf C}_{3}^{\rm r}(z)$ now generates basis functions of the form $Q_3(k_1-1,0,m)$, where $m$ are non-negative integers. These states are subsequently mapped back by ${\bf L}_{32}$ to basis functions $Q_2(k_1,m)$, which are no longer restricted to neighboring values of the second index. 
As a result, the memory kernel in the second equation loses its tridiagonal structure and instead acquires long-range couplings throughout the basis set. In short, the continued-fraction construction developed for the one- and two-domain truncations can no longer be applied, and the analytical solution breaks down at the three-domain level.

\subsection{Mode-Coupling approximations}
\label{sec:MCT}

Analytical studies based on the MCT constitute one of the earliest attempts to describe the dynamics of the East model. Although different closure schemes have been proposed over the years, they share the common philosophy of constructing self-consistent equations for the time correlation function.

The central assumption of MCT is that the irreducible memory kernel, defined in Eq.~(\ref{eq:relationofMandMir}), can be approximated as a functional of the correlation function.
In general, this functional is expanded to a  polynomial,
\begin{equation}
    M^{\rm ir}(t)=\sum_n{a_n[C(t)]^{n}}.
    \label{eq:ModeCouplingApprox}
\end{equation}

Kawasaki proposed the quadratic closure
\begin{equation}
    M^{\rm ir}(t) \approx c(1-c)[C(t)]^2,
    \label{eq:irrMofMCA}
\end{equation}
which is referred to as the mode-coupling approximation (MCA) because it is the standard quadratic closure widely used in conventional MCT~\cite{kawasaki1995irreducible}. 
Another commonly used closure is the effective-medium approximation (EMA) introduced by Eisinger and Jäckle~\cite{eisinger1993analytical},
\begin{equation}
    M^{\rm ir}(t)\approx c(1-c)C(t),
    \label{eq:irrMofEMA}
\end{equation}
which was later termed the simple mode-coupling (SMC) approximation by Wu and Cao \cite{wu2004east}. 

Despite their different closures, these approximations share similar qualitative behavior. They provide a reasonable description of the short-time or relatively high-temperature dynamics, but both predict a spurious ergodic-nonergodic transition at finite temperature. In particular, the MCA predicts a nonzero long-time limit of the correlation function for $c\leq 0.2$, whereas the EMA 
predicts ergodicity breaking already for $c\leq 0.5$. These finite-temperature transitions are inconsistent with Monte Carlo simulations and with rigorous spectral-gap analyses, both of which indicate that the East model remains ergodic for every finite excitation concentration $c>0$.

The breakdown of SMC approximation at low temperatures motivated several attempts to improve the mode-coupling description. One approach is the extended mode-coupling (EMC) approximation proposed by Pitts and Anderson \cite{pitts2001diagrammatic}, in which additional higher-order coupling terms are incorporated into the irreducible memory kernel.
Unlike the original SMC approximation, EMC removes the spurious ergodicity-breaking transition at a finite temperature. 
However,  its quantitative accuracy still deteriorates in the low-temperature regime ($c\lesssim0.5$), particularly for the long-time relaxation.

An alternative strategy is the long-time correction (LTC) introduced in Ref~\cite{wu2004east}.
Rather than modifying the mode-coupling functional itself, LTC is derived from the asymptotic long-time behavior of the irreducible memory kernel for finite chains. This phenomenological correction also eliminates the artificial ergodicity breaking produced by SMC and yields improved agreement with simulations.
However, despite these improvements, the convergence of the relaxation time with increasing approximation order remains unsatisfactory. Even when the long-time correction is carried to the fourth order, the predicted relaxation time still scales asymptotically as $\tau\sim 1/c$, indicating that increasing the truncation order does not alter the low-temperature scaling behavior. Consequently, the essential hierarchical activation mechanism responsible for the super-Arrhenius dynamics is still absent from the theory.

Overall, these approximations illustrate both the strengths and limitations of memory-kernel closures. They provide qualitatively reasonable description over a limited temperature range, but their predictive power at low temperatures depends sensitively on the particular closure adopted. Although various extensions alleviate the spurious ergodic-nonergodic transition, a systematic treatment of the increasing hierarchical relaxation dynamics remains challenging within this framework.

\section{Comparison with simulations}
\label{sec:simulations}

\begin{figure}
    \centering
    \includegraphics[width=0.9\linewidth]{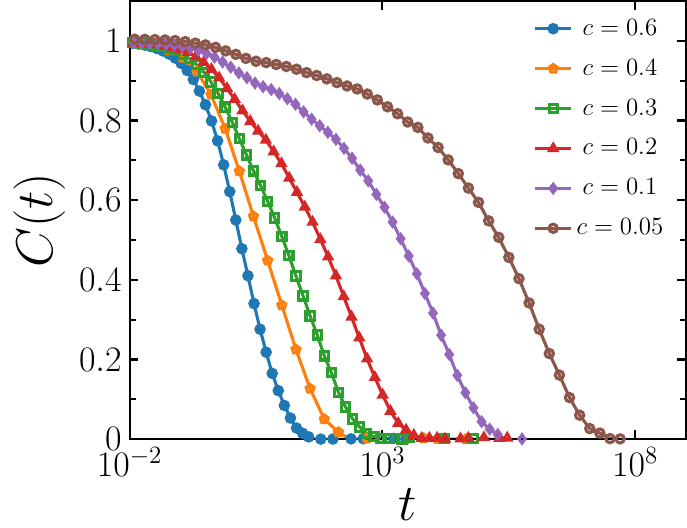}
    \caption{Continuous-time Monte Carlo simulation results for the correlation function $C(t)$ of the East model, obtained for a chain of length $N=328$ with periodic boundary conditions.}
    \label{fig:CtSimulation}
\end{figure}

This section examines the properties of the single-spin time correlation function $C(t)$ in the East model and compares the predictions of several existing theories against simulation results. As illustrated in Fig.~\ref{fig:CtSimulation}, continuous-time Monte Carlo simulations \cite{newman1999monte} show that, upon lowering $c$ (or, equivalently, the temperature $T$), the relaxation of the correlation function evolves from a single-step process into a multistep one, accompanied by a rapid increase in the relaxation time.

\begin{figure}
    \centering
    \includegraphics[width=0.9\linewidth]{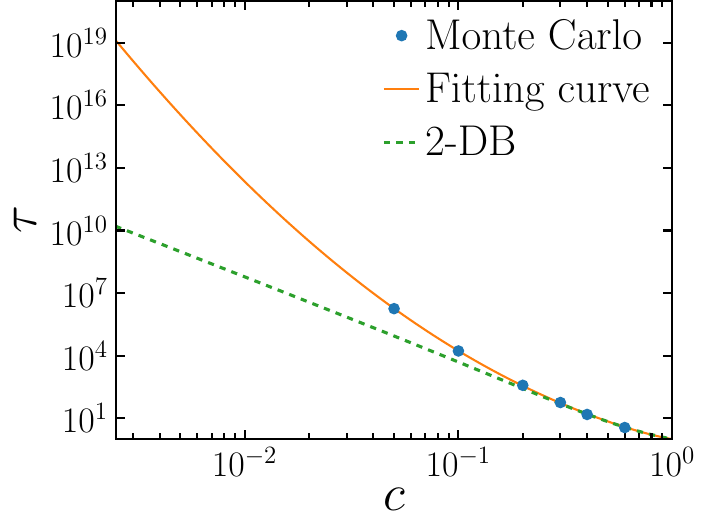}
    \caption{The relaxation time $\tau$ as a function of the mean occupancy $c$. The Monte Carlo simulation result is fitted by an empirical function $\frac{1}{c}\exp{\big[\frac{(\ln{c})^2}{2\ln{2}}+A\ln{\frac{1}{c}}+B\ln{(1+\ln{\frac{1}{c}})}+D(1-c)\big]}$, where $A\approx3.98$, $B\approx-9.04$ and $D\approx5.83$.
    This empirical form matches the exact scaling Eq.~(\ref{eq:tau}) in the limit $c \to 0$.
    }
    \label{fig:TauCompare}
\end{figure}

We first compare the relaxation times obtained from  different approaches. In the low-temperature limit $c\to 0$, the two-domain basis (2-DB) theory predicts a power-law divergence $\tau^{(2)}_{\rm DB}\sim 1/c^4$ \cite{van2005glassy}, 
whereas the exact result 
is $\tau\sim \exp{[\frac{(\ln{c})^2}{2\ln{2}}]}$ (see Eq.~\ref{eq:tau}), which diverges faster than any power law. The comparison of the relaxation times is shown in Fig.~\ref{fig:TauCompare}.
Note that the  fluctuation basis with mode-coupling closures predicts ergodicity breaking (a divergence of the relaxation time) at a finite temperature. This prediction is clearly inconsistent with the simulation results, which show that the relaxation time diverges only at zero temperature.

\begin{figure}
    \centering
    \includegraphics[width=0.9\linewidth]{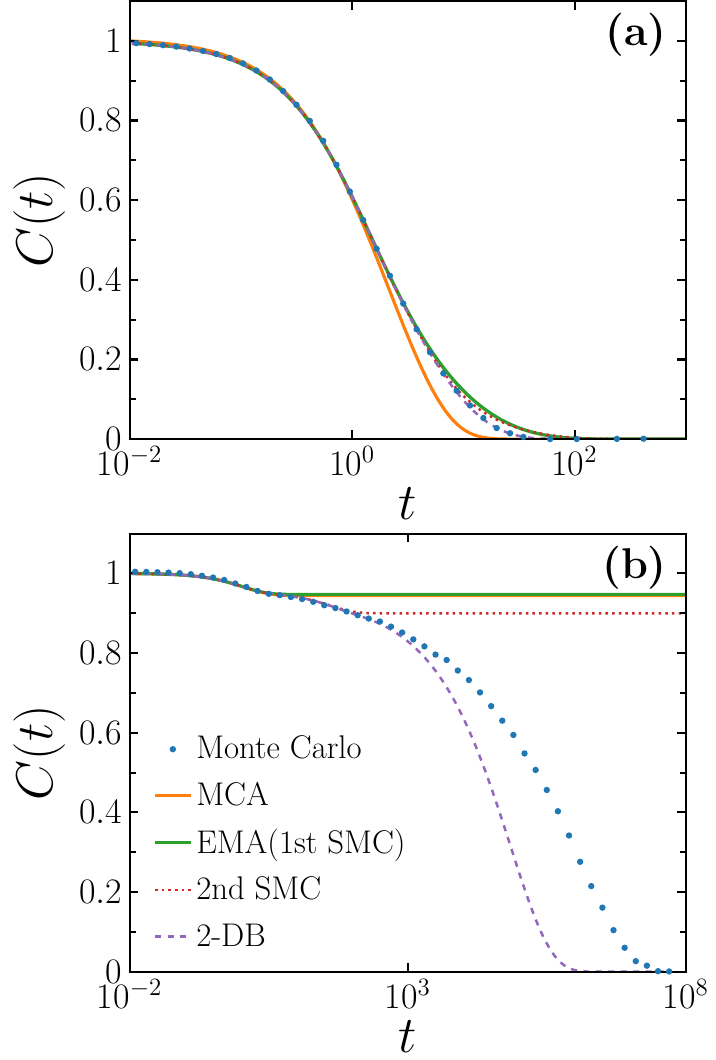}
    \caption{Comparison of the correlation function $C(t)$ between simulations and theoretical predictions at {(a)} $c=0.6$ and {(b)} $c=0.05$.}
    \label{fig:CtCompare}
\end{figure}

Figure~\ref{fig:CtCompare} compares the simulated $C(t)$ with that obtained from four existing theoretical approaches: (i) the mode-coupling approximation (MCA) \cite{kawasaki1995irreducible}, (ii) the effective-medium approximation (EMA) \cite{eisinger1993analytical}, or equivalently, the first-order simple mode-coupling (1st SMC) closure for the fluctuation basis \cite{wu2004east}, (iii) the second-order simple mode-coupling (2nd SMC) closure for the fluctuation basis \cite{wu2004east}, and (iv) the two-domain basis (2-DB) \cite{van2005glassy}. 
At high mean occupancy ($c=0.6$), all these theories give results that are close to the simulation data (Fig.~\ref{fig:CtCompare}a). As the temperature is lowered, however, different approximations successively begin to fail. The EMA (equivalent to 1st SMC) and the 2nd SMC exhibit an ergodic-nonergodic transition at $c=0.5$ \cite{wu2004east}. 
This is in contrast to the exact behavior of the East model, which is ergodic down to $c=0$ ($T_{\rm c}=0$).
As shown in Fig.~\ref{fig:CtCompare}b, the simulated $C(t)$ asymptotically decays to zero at $c=0.05$, indicating ergodic behavior. Similarly, the MCA fails at low $c$ by predicting a nonexistent ergodic-nonergodic transition at $c=0.2$ \cite{kawasaki1995irreducible}.

In contrast to the other three theories, the two-domain basis theory predicts that $C(t)$ remains ergodic down to zero temperature. However, its quantitative accuracy gradually deteriorates below $c\simeq 0.3$ \cite{van2005glassy}. This is a consequence of the truncation, which fails to retain all slow dynamical modes. In particular, for $c<0.3$, the predicted long-time relaxation is faster than that observed in the Monte Carlo simulations, as illustrated in Fig.~\ref{fig:CtCompare}b.

\section{Hierarchical relaxation in microscopic dynamics}
\label{sec:hierarchical}

\begin{figure}
    \centering
    \includegraphics[width=\linewidth]{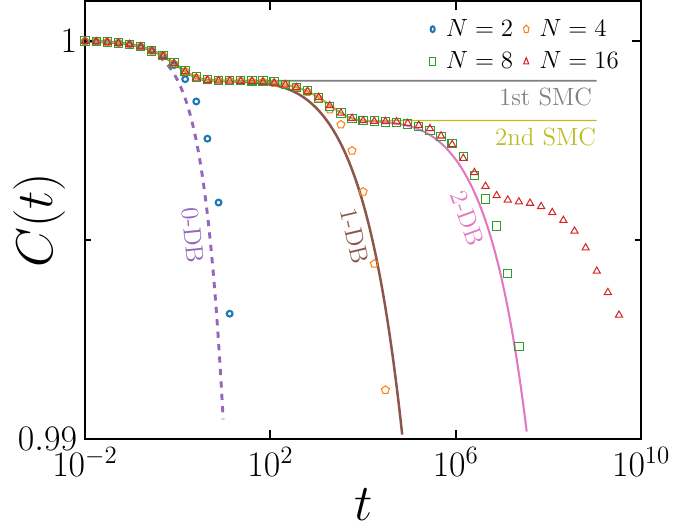}
    \caption{Microscopic hierarchical relaxation of the correlation function $C(t)$. Symbols represent simulation results for different system sizes, all obtained with free boundary conditions, where the rightmost spin is permanently fixed in the facilitated state. Solid lines show analytical predictions from different theoretical approaches for comparison. All results are obtained at $c=0.001$.}
    \label{fig:CtHierarchical}
\end{figure}

As the temperature \(T\) decreases, the relaxation time \(\tau\) of the East model grows rapidly. However, this growth does not involve a uniform evolution of relaxation time scales. Instead, at low \(T\) the dynamics exhibits a pronounced hierarchical structure, in which successive relaxation processes are separated by increasingly large time scales (see Fig.~\ref{fig:CtHierarchical}). The physical origin of this behavior can be understood in terms of the relaxation of domains.

According to Eq.~(\ref{eq:tau_total}), the relaxation time depends on the size of the domain under consideration. On the equilibrium time scale \(\tau\) (see Eq.~\ref{eq:tau}), all allowed domains are sampled. On short time scales, only small domains can relax. Based on these considerations we can divide the dynamical behavior of \(C(t)\) into three regimes.

{\it (i) Microscopic dynamical regime with hierarchical relaxation}: \(d \sim O(1) \ll 1/c \approx e^{1/T}\) and \(T \log t \sim O(1)\). In this regime, only domains of size \(O(1)\) are relaxed. Because \(d \sim O(1)\), both the barrier height and the entropy are of order \(O(1)\): \(h(d) \sim \log_2 d \sim O(1)\) (see Eq.~\ref{eq:domainhierarchy}) and \(s(d) \sim \log 2^{h^2/2} \sim O(1)\) (see Eq.~\ref{eq:entropy}). In the limit \(T \to 0\), the term \(T s(d)\) in the effective free energy, Eq.~(\ref{eq:free-energy}), vanishes. Thus, in this short-time regime, entropic effects are negligible, and the dynamics is dominated by the hierarchical structure of the barrier height, Eq.~(\ref{eq:domainhierarchy}).
Equation~(\ref{eq:tau_total}) shows that the time scale corresponding to $d \sim O(1)$ is $T \log t \sim O(1)$.

To see why the hierarchical barriers give rise to hierarchical relaxation dynamics, we note that Eq.~(\ref{eq:domainhierarchy}) naturally suggests a set of ``energy levels'' \(h \sim n\) and corresponding domain sizes \(d \sim 2^n\), where \(n = 0, 1, \ldots\).
The relaxation time of the \(n\)-th level is (see Eq.~\ref{eq:tau_energy_relation})
\begin{equation}
    \tau_n \sim e^{n/T} \approx c^{-n}.
    \label{eq:tau_n}
\end{equation}
This leads to a property known as \textit{time-scale separation}~\cite{chleboun2014time}.
The time axis can be divided into successive intervals,
\(
    (1,c^{-1}),\quad (c^{-1},c^{-2}),\quad (c^{-2},c^{-3}),\quad \dots
\)
Within the interval \((c^{-(n-1)},c^{-n})\), the relaxation is predominantly associated with domains of lengths \(2^{n-1} < d \leq 2^n\) and barriers of height \(h \sim n\). The relaxation time scale in this interval is \(\tau_n\), given by Eq.~(\ref{eq:tau_n}).

\textit{(ii) Mesoscopic dynamical regime}: 
$d=O(1/c^{\gamma})$ with $0<\gamma<1$, the relaxation dynamics occurs on time scales intermediate between the microscopic and equilibrium regimes. Rigorous results establish the persistence of time-scale separation at such mesoscopic lengths~\cite{chleboun2014time}.

To estimate the relaxation time $t_\gamma$, we consider the competition between energetic barriers and entropic contributions. The characteristic barrier height scales as $h(d)\sim \log_2d\sim \frac{\gamma}{T\ln 2}$, while the configurational entropy associated with the relaxation pathways scales as $s(d)\sim \frac{h(d)^2\ln 2}{2}$. Substituting these estimates into the effective free-energy barrier, Eq.~\eqref{eq:free-energy}, gives
$T^2\ln{t}\sim \frac{f(\gamma)}{2\ln{2}} = \frac{\gamma(2-\gamma)}{2\ln{2}}$, for 
    $\quad 0<\gamma<1$.
This expression provides a phenomenological interpolation between the microscopic and equilibrium relaxation scales. In particular, the crossover function $f(\gamma)=\gamma(2-\gamma)$ increases monotonically from $f(0)=0$ to $f(1)=1$, recovering the leading equilibrium relaxation-time scaling as $\gamma\to1$.



Here, the entropic effects become increasingly important. Nevertheless, rigorous results on time-scale separation show that the hierarchical structure persists on such mesoscopic length scales~\cite{chleboun2014time}. Although the growing number of accessible configurations affects the relaxation dynamics, it does not eliminate the separation between distinct barrier levels.

{\it (iii) Macroscopic (equilibrium) dynamical regime}: $d=O(1/c)$ and \(T^2 \log t \sim \frac{1}{2 \log 2}\). On this time scale, entropic contributions become important enough to compete with the energetic hierarchy, and the sharp distinction between successive relaxation levels disappears. Time-scale separation breaks down, and the hierarchical dynamics is no longer observable.

Next, we illustrate the short-time (regime (i)) hierarchical relaxation behavior using simulation results for \(C(t)\) with \(c=0.001\) in Fig.~\ref{fig:CtHierarchical}. The data do not capture the mesoscopic and macroscopic dynamics (regimes (ii) and (iii)), because reaching such long time scales in large systems is beyond current computational capabilities at such low temperatures. Our key intention is to demonstrate the hierarchical microscopic dynamics, as reflected by the multiple  plateaus in \(C(t)\) in Fig.~\ref{fig:CtHierarchical}. Note that these multiple plateaus are visible only at sufficiently low temperatures.

To demonstrate that the observed multiple plateaus in \(C(t)\) indeed originate from the relaxation of domains at different scales, we simulate small systems with a total of \(N = 2^n\) spins, with $n=1,2,3,4$. For \(c=0.001\), the equilibrium characteristic domain length is \(\langle d \rangle \sim 1/c = 1000\) (see Eq.~\ref{eq:domainLengthEq}), which is much larger than the simulated system size \(N\). This ensures that the dynamics remains in the microscopic regime (i).
For a chain of length \(N=2\), a simple exponential decay occurs, with a characteristic relaxation time \(\tau_1 \sim c^{-1} = 10^3\).
For \(N=4\), the system can contain domains extending to the next barrier level, \(n=2\), resulting in a secondary exponential decay with \(\tau_2 \sim c^{-2} = 10^6\); this gives rise to an apparent additional plateau. This procedure extends up to the largest simulated system size, \(N = 2^4 = 16\), confirming that domain relaxation is the origin of the hierarchical dynamics.

Remarkably, this hierarchical structure is also reflected in the successive approximations developed within the Zwanzig--Mori framework, using both fluctuation and domain bases (see Secs.~\ref{sec:fluctuation} and~\ref{sec:domain}). We first discuss the domain basis.
In Fig.~\ref{fig:CtHierarchical}, the zeroth-order domain-basis (0-DB) approximation corresponds to truncating Eq.~\eqref{eq:k-thEq} at \(k=0\), or equivalently setting the memory kernel \(\tilde{M}_0(z)=0\). This yields simple exponential relaxation, \(C(t)=e^{-ct}\), whose characteristic time scale is \(\tau_1 \sim c^{-1}\). It represents the fastest relaxation mode, which is also present in the higher-order approximations. The 0-DB approximation yields the first plateau. The first-order domain-basis (1-DB) approximation retains this fastest mode with time scale \(\tau_1\) and, as seen in Fig.~\ref{fig:CtHierarchical}, resolves the second plateau. However, the late-time relaxation after this plateau is governed by a characteristic scale of \(\tau_2 \sim c^{-3}\)~\cite{van2005glassy}, rather than the naive domain-level estimate $c^{-2}$. By including one more basis set, the second-order domain-basis (2-DB) approximation resolves the third plateau and extends the late-time description to scale \(\tau_3 \sim c^{-4}\)~\cite{van2005glassy}. Thus, the order of the domain-basis truncation should be associated with plateau it resolves, not with the domain level $n$ in $\tau_n\sim c^{-n}$. The extra factor of $c^{-1}$ relative to the naive domain-level estimate appears in the relaxation after the last resolved plateau, reflecting the additional correction introduced by the domain-basis closure. Unfortunately, as discussed above, the analytic method used to obtain these results becomes intractable at third order, preventing a corresponding analytical description of the next relaxation level.


A similar hierarchy appears in the successive approximations of the SMC approach using the fluctuation basis. In the first-order SMC (1st SMC) approximation, the closure assumption captures the second plateau at \(C_2=(1-2c)/(1-c)\), but fails to describe the subsequent decay toward zero (note that \(C_1 = 1\)). The second-order SMC (2nd SMC) approximation extends the description to the next plateau at \(C_3 = 1-2c\), but likewise fails to recover the complete long-time relaxation. Moreover, the structure of the closure scheme prevents a straightforward extension to higher orders.

These results illustrate both the ability of low-order theoretical approximations to capture the initial levels of hierarchical relaxation and the difficulty of systematically incorporating the increasingly broad range of time scales. Interestingly, the lower-order approximations in both the DB and SMC schemes correctly capture \(C(t)\) for small systems at the corresponding size levels. For larger systems, however, they fail in different ways: the closure scheme in SMC produces unphysical plateaus, whereas the truncation scheme in DB neglects slower modes.

\section{Discussion}
\label{sec:discussion}

Despite being one of the simplest kinetically constrained models, the East model still lacks a rigorous description of its time correlation function. 
As discussed above, all existing theoretical approaches eventually fail at sufficiently low temperatures. 
Several commonly used closure schemes predict spurious dynamical critical points, while truncation-based approaches fail to reproduce the correct long-time behavior and the asymptotic divergence of the relaxation time at low temperatures.

The underlying difficulty originates from the highly hierarchical nature of the East model dynamics. As the hierarchy is extended, the complexity of the corresponding dynamical problem increases dramatically. More importantly, the increase in complexity is accompanied by a rapid loss of the algebraic structures that make the hierarchy analytically tractable. This is illustrated by the difficulties encountered beyond the second order in both the fluctuation-basis and domain-basis approaches. Thus, the central challenge is not simply the increasing number of dynamical modes, but the fact that the analytically tractable structure itself becomes progressively more complicated as higher levels of the hierarchy are included.

The existing approaches based on the Zwanzig-Mori formalism can be broadly divided into closure-based and truncation-based strategies. Closure schemes assume that higher-order memory kernels can be expressed in terms of lower-order kernels. Such approximations can be remarkably effective at large $c$, but may introduce qualitatively incorrect long-time dynamics, including spurious ergodic-nonergodic transitions. Truncation-based approaches, in contrast, preserve the exact dynamics within a finite (or infinite) subspace and therefore avoid such artificial closures, but necessarily discard higher-order modes. Since increasingly slow modes become important at low temperatures, this leads to an incorrect long-time relaxation and, eventually, an incorrect asymptotic scaling of the relaxation time.

These observations suggest that a successful theoretical description of the low-temperature dynamics should retain some information about the hierarchical organization of relaxation processes, while avoiding the rapidly growing complexity of the full hierarchy. In this sense, the results discussed here highlight both the strength and the limitation of the existing projection-operator approaches: they provide a systematic framework for organizing the dynamics into successive levels, but finding an analytically tractable representation of the resulting hierarchy remains the central unresolved problem. 

In the second paper of this series~\cite{Zha2026Dynamical2}, we establish an alternative theoretical framework called the path-probability formalism. This theory yields an infinite but closed set of equations for single-spin path-probability densities. Although exact solutions of these equations remain out of reach, a systematic factorization approximation scheme is discussed.

\section{Acknowledgements}
We thank Chengjie Luo and Haijun Zhou for useful discussions. We acknowledge funding from the National Key R\&D Program of China (Grant No.~2025YFF0512000), the China Manned Space Program (Grant No.~CMSS-2025-5-P-002), the Wenzhou Institute (No.~WIUCASICTP2022), and the National Natural Science Foundation of China (Nos.~12447101).
We acknowledge the use of the High-Performance Cluster at the Institute of Theoretical Physics, Chinese Academy of Sciences.

\bibliography{EastRef}

@article{parisi1979infinite,
  title={Infinite Number of Order Parameters for Spin-Glasses},
  author={Parisi, Giorgio},
  journal={Physical Review Letters},
  volume={43},
  number={23},
  pages={1754--1756},
  year={1979},
  doi={10.1103/PhysRevLett.43.1754}
}

@article{parisi1980order,
  title={The Order Parameter for Spin Glasses: A Function on the Interval 0-1},
  author={Parisi, Giorgio},
  journal={Journal of Physics A: Mathematical and General},
  volume={13},
  number={3},
  pages={1101--1101},
  year={1980},
  doi={10.1088/0305-4470/13/3/042}
}

@article{binder1986spin,
  title={Spin glasses: Experimental facts, theoretical concepts, and open questions},
  author={Binder, Kurt and Young, A. P.},
  journal={Reviews of Modern Physics},
  volume={58},
  pages={801--976},
  year={1986},
  doi={10.1103/RevModPhys.58.801}
}

@article{cipelletti2005slow,
  title={Slow dynamics in glassy soft matter},
  author={Cipelletti, Luca and Ramos, Laurence},
  journal={Journal of Physics: Condensed Matter},
  volume={17},
  pages={R253--R285},
  year={2005},
  doi={10.1088/0953-8984/17/6/R01}
}

@article{berthier2011theoretical,
  title={Theoretical perspective on the glass transition and amorphous materials},
  author={Berthier, Ludovic and Biroli, Giulio},
  journal={Reviews of Modern Physics},
  volume={83},
  pages={587--645},
  year={2011},
  doi={10.1103/RevModPhys.83.587}
}

@article{zwanzig1960ensemble,
  title={Ensemble method in the theory of irreversibility},
  author={Zwanzig, Robert},
  journal={The Journal of Chemical Physics},
  volume={33},
  number={5},
  pages={1338--1341},
  year={1960},
  publisher={American Institute of Physics}
}

@article{mori1965transport,
  title={Transport, collective motion, and Brownian motion},
  author={Mori, Hazime},
  journal={Progress of theoretical physics},
  volume={33},
  number={3},
  pages={423--455},
  year={1965},
  publisher={Oxford University Press}
}

@article{kawasaki1995irreducible,
  title={Irreducible memory function for dissipative stochastic systems with detailed balance},
  author={Kawasaki, Kyozi},
  journal={Physica A: Statistical Mechanics and its Applications},
  volume={215},
  number={1-2},
  pages={61--74},
  year={1995},
  publisher={Elsevier}
}

@article{jackle1991hierarchically,
  title={A hierarchically constrained kinetic Ising model},
  author={J{\"a}ckle, Josef and Eisinger, Siegfried},
  journal={Zeitschrift f{\"u}r physik B condensed matter},
  volume={84},
  number={1},
  pages={115--124},
  year={1991},
  publisher={Springer}
}

@article{eisinger1993analytical,
  title={Analytical approximations for the hierarchically constrained kinetic Ising chain},
  author={Eisinger, S and J{\"a}ckle, J},
  journal={Journal of statistical physics},
  volume={73},
  number={3},
  pages={643--670},
  year={1993},
  publisher={Springer}
}

@article{sollich1999glassy,
  title={Glassy time-scale divergence and anomalous coarsening in a kinetically constrained spin chain},
  author={Sollich, Peter and Evans, Martin R},
  journal={Physical review letters},
  volume={83},
  number={16},
  pages={3238},
  year={1999},
  publisher={APS}
}

@book{newman1999monte,
  title={Monte Carlo Methods in Statistical Physics},
  author={Newman, M.E.J. and Barkema, G.T.},
  isbn={9780198517979},
  lccn={99213405},
  url={https://books.google.com.hk/books?id=J5aLdDN4uFwC},
  year={1999},
  publisher={Clarendon Press}
}

@article{pitts2001diagrammatic,
  title={A diagrammatic theory of time correlation functions of facilitated kinetic Ising models},
  author={Pitts, Steven J and Andersen, Hans C},
  journal={The Journal of Chemical Physics},
  volume={114},
  number={3},
  pages={1101--1114},
  year={2001},
  publisher={American Institute of Physics}
}

@article{chung2001combinatorics, author = {Chung, Fan and Diaconis, Persi and Graham, Ronald}, title = {Combinatorics for the East Model}, year = {2001}, issue_date = {July 2001}, publisher = {Academic Press, Inc.}, address = {USA}, volume = {27}, number = {1}, issn = {0196-8858}, url = {https://doi.org/10.1006/aama.2001.0728}, doi = {10.1006/aama.2001.0728}, journal = {Adv. Appl. Math.}, month = jul, pages = {192–206}, numpages = {15} }

@article{aldous2002asymmetric,
  title={The asymmetric one-dimensional constrained Ising model: rigorous results},
  author={Aldous, David and Diaconis, Persi},
  journal={Journal of statistical physics},
  volume={107},
  number={5},
  pages={945--975},
  year={2002},
  publisher={Springer}
}

@article{Ritort2003Glassy,
author = {Ritort, F. and Sollich, Peter},
year = {2003},
month = {06},
pages = {219-342},
title = {Glassy dynamics of kinetically constrained models},
volume = {52},
journal = {Advances in Physics},
doi = {10.1080/0001873031000093582}
}

@article{cancrini2008kinetically,
  title={Kinetically constrained spin models},
  author={Cancrini, Nicoletta and Martinelli, Fabio and Roberto, Cyril and Toninelli, Cristina},
  journal={Probability theory and related fields},
  volume={140},
  number={3},
  pages={459--504},
  year={2008},
  publisher={Springer}
}

@article{toninelli2012aging,
  title={Aging through hierarchical coalescence in the East model},
  author={Toninelli, C and Faggionato, A and Martinelli, F and Roberto, C},
  journal={HAL},
  volume={2012},
  year={2012}
}

@article{sollich2003glassy,
  title={Glassy dynamics in the asymmetrically constrained kinetic Ising chain},
  author={Sollich, Peter and Evans, MR},
  journal={Physical Review E},
  volume={68},
  number={3},
  pages={031504},
  year={2003},
  publisher={APS}
}

@article{wu2004east,
  title={East model: Basis set expansion, mode coupling, and irreducible memory kernels},
  author={Wu, Jianlan and Cao, Jianshu},
  journal={The Journal of Physical Chemistry B},
  volume={108},
  number={21},
  pages={6796--6808},
  year={2004},
  publisher={ACS Publications}
}

@article{van2005glassy,
  title={Glassy dynamics and domains: Explicit results for the East model},
  author={van Zon, Ramses and Schofield, Jeremy},
  journal={The Journal of chemical physics},
  volume={122},
  number={19},
  year={2005},
  publisher={AIP Publishing}
}

@article{cancrini2010facilitated,
  title={Facilitated oriented spin models: some non equilibrium results},
  author={Cancrini, Nicoletta and Martinelli, Fabio and Schonmann, Roberto and Toninelli, Cristina},
  journal={Journal of statistical physics},
  volume={138},
  number={6},
  pages={1109--1123},
  year={2010},
  publisher={Springer}
}

@article{chleboun2014time,
  title={Time scale separation and dynamic heterogeneity in the low temperature East model},
  author={Chleboun, Paul and Faggionato, Alessandra and Martinelli, Fabio},
  journal={Communications in mathematical physics},
  volume={328},
  number={3},
  pages={955--993},
  year={2014},
  publisher={Springer}
}

@article{bouchaud1997aging,
  title={Aging in glasses: Traps and mode-coupling theory},
  author={Bouchaud, Jean-Philippe and M{\'e}zard, Marc},
  journal={Progress of Theoretical Physics Supplement},
  volume={126},
  pages={181--184},
  year={1997},
  publisher={Oxford Academic}
}

@article{li2026universal,
  title={Universal activated aging and weak ergodicity breaking in spin and structural glasses},
  author={Li, Bin and Pan, Deng and Qu, Ting and Jin, Yuliang},
  journal={Science Advances},
  volume={12},
  number={9},
  pages={eaec4416},
  year={2026},
  publisher={American Association for the Advancement of Science}
}

@article{pitts2000facilitated,
  title={Facilitated spin models, mode coupling theory, and ergodic--nonergodic transitions},
  author={Pitts, Steven J and Young, Thomas and Andersen, Hans C},
  journal={The Journal of Chemical Physics},
  volume={113},
  number={19},
  pages={8671--8679},
  year={2000},
  publisher={American Institute of Physics}
}

@article{biezemans2020glassy,
  title={Glassy dynamics from generalized mode-coupling theory: Existence and uniqueness of solutions for hierarchically coupled integro-differential equations},
  author={Biezemans, Rutger A and Ciarella, Simone and {\c{C}}aylak, Onur and Baumeier, Bj{\"o}rn and Janssen, Liesbeth MC},
  journal={Journal of Statistical Mechanics: Theory and Experiment},
  volume={2020},
  number={10},
  pages={103301},
  year={2020},
  publisher={IOP Publishing and SISSA}
}

@article{luo2020generalized1,
  title={Generalized mode-coupling theory of the glass transition. I. Numerical results for Percus--Yevick hard spheres},
  author={Luo, Chengjie and Janssen, Liesbeth},
  journal={The Journal of Chemical Physics},
  volume={153},
  number={21},
  year={2020},
  publisher={AIP Publishing}
}

@article{luo2020generalized2,
  title={Generalized mode-coupling theory of the glass transition. II. Analytical scaling laws},
  author={Luo, Chengjie and Janssen, Liesbeth},
  journal={The Journal of Chemical Physics},
  volume={153},
  number={21},
  year={2020},
  publisher={AIP Publishing}
}

@article{janssen2016generalized,
  title={Generalized mode-coupling theory of the glass transition: schematic results at finite and infinite order},
  author={Janssen, Liesbeth MC and Mayer, Peter and Reichman, David R},
  journal={Journal of Statistical Mechanics: Theory and Experiment},
  volume={2016},
  number={5},
  pages={054049},
  year={2016},
  publisher={IOP Publishing and SISSA}
}

@article{mayer2006cooperativity,
  title={Cooperativity beyond caging: Generalized mode-coupling theory},
  author={Mayer, Peter and Miyazaki, Kunimasa and Reichman, David R},
  journal={Physical review letters},
  volume={97},
  number={9},
  pages={095702},
  year={2006},
  publisher={APS}
}

@article{bouchaud1995aging,
  title={Aging on Parisi's tree},
  author={Bouchaud, J-P and Dean, David S},
  journal={Journal de Physique I},
  volume={5},
  number={3},
  pages={265--286},
  year={1995}
}

@article{cugliandolo1993analytical,
  title={Analytical solution of the off-equilibrium dynamics of a long-range spin-glass model},
  author={Cugliandolo, Leticia F and Kurchan, Jorge},
  journal={Physical review letters},
  volume={71},
  number={1},
  pages={173},
  year={1993},
  publisher={APS}
}

@article{angell1995formation,
  title={Formation of glasses from liquids and biopolymers},
  author={Angell, C Austen},
  journal={Science},
  volume={267},
  number={5206},
  pages={1924--1935},
  year={1995},
  publisher={American Association for the Advancement of Science}
}

@article{dahlberg2025spin,
  title={Spin-glass dynamics: Experiment, theory, and simulation},
  author={Dahlberg, ED and Gonz{\'a}lez-Adalid Pemart{\'\i}n, I and Marinari, E and Parisi, G and Ricci-Tersenghi, F and Martin-Mayor, V and Moreno-Gordo, J and Orbach, RL and Paga, I and Ruiz-Lorenzo, JJ and others},
  journal={Reviews of Modern Physics},
  volume={97},
  number={4},
  pages={045005},
  year={2025},
  publisher={APS}
}

@article{ikeda2012unified,
  title={Unified study of glass and jamming rheology in soft particle systems},
  author={Ikeda, Atsushi and Berthier, Ludovic and Sollich, Peter},
  journal={Physical review letters},
  volume={109},
  number={1},
  pages={018301},
  year={2012},
  publisher={APS}
}

@article{Zha2026Dynamical2,
  title={Dynamical theory of the East model. II. A path-probability formalism},
  author={Zha, Chong and Jin, Yuliang},
  journal={arXiv:2611.23008},
  volume={},
  number={},
  pages={},
  year={2026}
}
\end{document}